\documentclass{article}
\usepackage{url,framed}
\usepackage[nosumlimits]{amsmath}
\usepackage{amssymb,amsthm,MnSymbol}

\newcommand{\dd}[2]{\frac{\diff#1}{\diff#2}}

\usepackage{amscd}

\usepackage[margin=2cm]{geometry}

\def\dd{{\color{red}\mathsf d}}

\usepackage{fancybox}

\usepackage[color=yellow]{todonotes}
\usepackage[colorlinks]{hyperref}
\hypersetup{colorlinks,linkcolor={red},citecolor={blue},urlcolor={green!70!black}}

\begin{document}

\title{Predicting Uncertainty in Geometric Fluid Mechanics}

\author{Fran\c{c}ois Gay-Balmaz$^1$ and Darryl D Holm$^2$
}
\addtocounter{footnote}{1}
\footnotetext{CNRS and \'Ecole Normale Sup\'erieure de Paris, Laboratoire de M\'et\'eorologie Dynamique, 24 Rue Lhomond, 75005 Paris, France. 
\texttt{gaybalma@lmd.ens.fr}
\addtocounter{footnote}{1} }
\footnotetext{Department of Mathematics, Imperial College, London SW7 2AZ, UK. 
\texttt{d.holm@ic.ac.uk}
\addtocounter{footnote}{1}}

\date{\it In Honour of Jeurgen Scheurle's 65th, Happy Birthday!}

\maketitle

\makeatother

\begin{abstract}
We review opportunities for stochastic geometric mechanics to incorporate observed data into variational principles, in order to derive data-driven nonlinear dynamical models of effects on the variability of computationally resolvable scales of fluid motion, due to unresolvable, small, rapid scales of fluid motion.
\end{abstract}

\tableofcontents

\section{Motivation}

Sensitive dependence on initial conditions limits deterministic prediction in nonlinear dynamics, because every physical measurement has uncertainty. That is, every measurement in Nature is stated with a probability of unknown error; since no two measurements of any physical quantity are ever exactly the same. In classical measurements, the effects of loss of predictability due to incomplete knowledge can sometimes be modelled as noise.  Noise can be used to model inaccuracy due to unknown errors in either measurement or computational simulation. In turn, the effects of the noise on the individual realisations of the solutions of dynamical systems can be associated under certain conditions with stochastic processes corresponding to a probability distribution.

In some physical situations, the intrinsically probabilistic nature and resulting inaccuracy in prediction is painfully clear. For example, the variability of the weather in many places on Earth has increased so much during the past century that rare extreme events which were once expected to happen every thousand years have recently been happening every year. This means that the old probability distributions for the variability of the weather as obtained from previous data no longer apply. Although the human-produced emissions of greenhouse gases are generally agreed among scientists to be the primary cause of this increased variability, many as yet unknown nonlinear consequences from it could still emerge that could increase the variability even further. Apparently, many ``unknown unknowns" are afoot. How shall we predict the uncertainty that they can produce? Because this imperative issue for humanity involves fluid dynamics of the atmosphere and ocean that is being continually observed, measured and simulated computationally, and because achieving the goal of predicting uncertainty and variability of the weather and climate must address the dynamics of fluids at the most basic level, we will start with the question raised by the following situation for atmospheric and oceanic observations.

Atmospheric and oceanic observations generally produce high quality data; but it is sparse. That is, the data has many gaps where information is missing. Moreover, the data is relatively local in both space and time. It is generally taken at scales in length and time that are far too small to be resolvable in computational simulations performed at the regional or even global scales which influence each other and that are necessary for the prediction of weather and climate events. For example, satellite observations produce immense amounts of high quality data along the paths of the satellite orbits, but not elsewhere at the same time. To be useful for input to computational simulations and verification of output from these simulations, the observed data that numerical computations of weather and climate \textit{cannot resolve} well enough to simulate in real time must generally be   interpolated, extrapolated and spread over scales that allow real-time computational simulations.  This process of ``upscaling", or ``coarse graining'' the data for use in computational simulations relies on data assimilation, which is essentially statistical. 

The question we must address in this situation is the following. \vspace{-3mm}
\begin{quote}
How can we use observed data in combination with the mathematics of stochastic processes in nonlinear dynamical systems to estimate and model those effects on the variability of computationally resolvable scales of motion that are caused by the small, rapid, unresolvable scales of fluid motion that upscaling in data assimilation leaves out?
\end{quote}


Recently, a partial answer to this question has been found in \cite{CoGoHo2017} by showing that a multi-scale decomposition of the deterministic Lagrange-to-Euler fluid flow map $g_t$ into a slow large-scale mean and a rapidly fluctuating small-scale map leads to Lagrangian fluid paths $x_t=g_tX$ with $g_0=Id$ on a manifold $ \mathcal{D} $ governed by the stochastic process $g_t\in {\rm Diff}(\mathcal{D})$ on the Lie group of diffeomorphic flows, which appears in the same form as had been proposed and studied for fluids in \cite{holm2015variational}; namely,
\begin{equation}
{\dd}x_t  = {\dd}g_t \,X
= u(x,t)dt + \sum_{i=1}^N \xi_i(x)\circ dW^i_t  
=   u(g_tX,t)dt + \sum_{i=1}^N \xi_i(g_tX)\circ dW^i_t 
\,,\label{Lag-stoch-process}
\end{equation}
where $x=g_tX$, ${\dd}$ represents stochastic time evolution, the vector fields $\xi_i(x)$ for $i=1,2,\dots,N,$ are prescribed functions of the Eulerian spatial coordinates, $x\in \mathcal{D}$ on the domain of flow $\mathcal{D}$, and $\circ\, dW^i(t)$ denotes the Stratonovich differential with independent Brownian motions $dW^i(t)$. The stochastic process for the evolution of the Lagrangian process $g_t$ in equation \eqref{Lag-stoch-process} involves the pullback $g_t^*$ of the Eulerian total velocity vector field, which comprises the sum of a drift displacement vector field $u_t(x)dt$ plus a sum over terms in $\xi_i(x)$ representing the (assumed stationary) spatial correlations of the temporal noise in the Stratonovich representation, each with its own independent Brownian motion in time.

The idea, then, is to regard the stochastic paths of certain \textit{tracers}, or advected quantities, $q(x,t)=q_0(X)g_t^{-1}\in Q$ for a manifold of variables $Q$ on which the Lagrangian stochastic process $g_t$ in \eqref{Lag-stoch-process} acts by smooth invertible maps, as observable data, e.g., from satellite observations. From this Lagrangian tracer data, one obtains the correlation eigenvectors $\xi_i(x)$ via the appropriate data assimilation methods, and uses these eigenvectors to derive the corresponding equations of motion for the smooth vector field $u_t\in \mathfrak{g}(\mathcal{D})$ by stochastically constraining Hamilton's variational principle for the fluid motion. Thereby, one obtains a variational approach for deriving \textit{data-driven} models in the framework of stochastic geometric mechanics via the Eulerian flow velocity decomposition in \eqref{Lag-stoch-process}. 

In \cite{holm2015variational} the velocity decomposition formula \eqref{Lag-stoch-process} was applied in the Hamilton-Clebsch variational principle to derive coadjoint motion equations as stochastic partial differential equations (SPDEs) whose ensemble of realisations  represented the uncertainty in the slow dynamics of the resolved mean velocity $u_t(x)$.  Under the conditions imposed in the derivation of formula \eqref{Lag-stoch-process} in \cite{CoGoHo2017} using homogenization theory, the sum of vector fields in \eqref{Lag-stoch-process} that had been treated in \cite{holm2015variational} from the viewpoint of stochastic coadjoint motion was found to represent a bona fide decomposition of the fluid transport velocity into a mean plus fluctuating flow. 

In what follows, we will discuss a few of the many opportunities for geometric mechanics to play a fundamental role in making the next advances in formulating, analysing and implementing stochastic fluid dynamics. In discussing these opportunities, we will describe approaches to stochastic variational principles which apply quite generally in geometric mechanics, so this introduction to stochastic geometric mechanics will not be limited to applications in fluid dynamics. For example, detailed expositions of the basic theory and applications of various approaches for a broad class of finite dimensional cases, see \cite{ArHoCa2017,CrHoRa2017}. 

\section{Structure preserving stochastic mechanics}\label{noise}

\subsection{Data-driven modelling of uncertainty} 

As opposed to theory-driven models such as Newtonian force laws and thermodynamic processes for the subgrid-scale dynamics, here we will introduce stochastic geometric mechanics as an opportunity to consider a stochastic version of data-driven modelling. In data-driven modelling, one seeks to model properties of a subsystem of a given dynamical system which, for example, may be observable at length or time scales which are below the resolution of available initial and boundary conditions, or scales finer than the resolution of numerical simulations of the dynamical system based on the assumed exact equations. 

The most familiar example of data-driven modelling occurs in numerical weather forecasting (NWF). In NWF, various numerically unresolvable, but observable, local subgrid-scale processes, such as formation of fronts and generation of tropical cyclones, are expected to have profound effects on the variability of the weather. These subgrid-scale processes must be parameterized at the resolved scales of the numerical simulations. Of course, the accuracy of a given parameterization model often remains uncertain. In fact, even the possibility of modelling subgrid-scale properties in terms of resolved-scale quantities available to simulations may sometimes be questionable. However, if some information about the \textit{statistics} of the small-scale excitations is known, such as the spatial correlations of its observed transport properties at the resolved scales, one may arguably consider modelling the effects of the small scale dynamics on the resolved scales by a stochastic transport process whose spatial correlations match the observations, at the computationally unresolvable scales. 
In this case, the eigenvectors of the correlation matrix of the observations may provide the modes of the subscale motion, to be modelled by applying stochasticity with the statistics of the unresolved scales.  Fluid dynamics is an ideal application for this approach; because it it falls within the purview of geometric mechanics \cite{arnold89mechanics}, and it has been a source of inspiration in the previous development of stochastic geometric mechanics \cite{holm2015variational}. 

\subsection{Geometric fluid dynamics}

Before inserting stochasticity, we first quickly review here the geometric formulation of ideal incompressible fluids. Following the geometric approach of \cite{Ar1966}, we consider the group $G=\operatorname{Diff}_{\rm vol}(\mathcal{D})$ of volume preserving diffeomorphisms of the fluid domain $ \mathcal{D} $, as the configuration manifold for incompressible fluids. The multiplication is given by composition of diffeomorphisms and we shall denote it by $g\circ h= gh$. Curves $g_t \in G$ in this group describe Lagrangian trajectories $x_t=g_t(X)$ of the fluid motion. To simplify our discussion, we will take $ \mathcal{D} $ as a bounded domain  in $ \mathbb{R}  ^2 $ or $ \mathbb{R}  ^3 $ with smooth boundary $\partial \mathcal{D}$. However, our developments extend easily to the case where $ \mathcal{D} $ is a Riemannian manifold with smooth boundary. Considering $G$ formally as a Lie group, its Lie algebra denoted $\mathfrak{g}$ is given by the space of divergence free vector fields on $\mathcal{D}$ parallel to the boundary $\partial\mathcal{D}$, endowed with the Lie bracket $[u,v]=v\cdot \nabla u - u\cdot\nabla v$.

The Lagrangian of the incompressible fluid is defined on the tangent bundle $TG$  of the group $G$ and is given by the kinetic energy, i.e., 
\begin{equation}\label{L_Euler}
L(g, v)=  \int _ \mathcal{D} \frac{1}{2}|v(X)| ^2 d ^{\,n} \!X,
\end{equation}
for $n=2,3$. For simplicity, we used the local notation $(g,v)$ for a vector in the tangent bundle $TG$ based at $g\in G$.
By a change of variables, we note that $L$ is right $G$-invariant, i.e., $L(gh, vh)=L(g, v)$, for all $h$ in $G$. Here again, we have used the simplified notation $(g,v)\mapsto (gh, vh)$ for the tangent lifted right action of $G$ on $TG$.

From its $G$-invariance, the Lagrangian $L$ yields the reduced Lagrangian $\ell: \mathfrak{g}  \rightarrow \mathbb{R}$ defined by $L(g, v)= \ell( v g ^{-1} )$. From expression \eqref{L_Euler}, we obtain
\begin{equation}\label{ell} 
\ell(u)= \int_ \mathcal{D}\frac{1}{2} | u(x) | ^2 d ^n x,
\end{equation} 
where  $u:=v g ^{-1}\in \mathfrak{g}$ is the Eulerian fluid velocity.

In the Lagrangian description, the equations of motion are evidently given by Hamilton's principle, written as
\begin{equation}\label{HP}
\delta \int_0^TL(g_t, \dot g_t) dt=0\,,
\end{equation}
for all variations of the curve $g_t$ with fixed extremities. The critical curves are solutions of the Euler-Lagrange equations for $L$ on $G$, which in the case of the Lagrangian \eqref{L_Euler} are geodesics of the weak $L^2$ Riemannian metric on $G$.

As a preparation for the stochastic extension below, we note that the dynamics in the Lagrangian description can be also obtained by the Hamilton-Pontryagin principle%
\footnote{This is a variational principle on the Pontryagin bundle $TG \oplus T^*G \rightarrow G$, defined as the vector bundle over $G$ with vector fiber at $g \in G$ given by $T_gG \oplus T^*_gG$.}
\begin{equation}\label{Ham_Pont}
\delta \int_0^T\left[L(g_t, v) + \left\langle \pi, \dot g_t- v\right\rangle \right]dt=0,
\end{equation}
for variations $\delta g_t$, $\delta v$, $\delta \pi$,  with $\delta g_t$ vanishing at $t=0,T$, see \cite{yoshimura2006dirac} and \cite{YoGB2011}. For simplicity, here and below, we shall indicate explicit time $t$ dependence only in the group variable $g_t$, with the understanding that $v$ and $\pi$ also depend on $t$. In this principle, the second order condition $\dot g_t=v$ is inserted as a constraint in the variational principle with the help of a Lagrange multiplier $\pi$ in the cotangent bundle $T^*G$ of $G$.
In \eqref{Ham_Pont} the angle brackets $ \left\langle\,\cdot \,, \,\cdot\,\right\rangle $ denote the duality pairing between elements in $T_g^\ast G$, and $T_g G$, the cotangent and tangent spaces of $G$ at $g$. Taking variations in the Hamilton-Pontryagin principle \eqref{Ham_Pont} yields, in local coordinates, the conditions
\begin{equation}\label{stat_cond_Ham_Pont}
\dot g_t= v,\qquad \frac{\partial L}{\partial v}=\pi,\qquad \frac{\partial L}{\partial g_t}=\dot \pi \,.
\end{equation}
These stationarity conditions identify $\pi$ as the material fluid momentum, and yield an implicit version of the Euler-Lagrange equations. The intrinsic expression of the stationarity conditions \eqref{stat_cond_Ham_Pont} can be given with the help of covariant derivatives. From the $G$-invariance of $L$, one obtains the Eulerian version of \eqref{Ham_Pont},
\begin{equation}\label{Ham_Pont_Eulerian}
\delta \int_0^T\left[\ell(u) + \left\langle m, \dot g_tg_t^{-1}- u\right\rangle_\mathfrak{g} \right]dt=0,
\end{equation}
for variations $\delta u$, $\delta g_t$ and $\delta m$, with $\delta g_t$ vanishing at $t=0,T$, and where $\left\langle\,\cdot \,, \,\cdot\,\right\rangle_\mathfrak{g}$ denotes the duality pairing between the Lie algebra $\mathfrak{g}$ and its dual $\mathfrak{g}^*$. This principle yields the conditions
\begin{equation}\label{Stat_cond_Eulerian}
\dot g_t g_t^{-1}= u,\qquad \frac{\delta \ell}{\delta u}=m,\qquad \partial_tm+ \operatorname{ad}^*_{\dot g_t g_t^{-1}}m=0,
\end{equation}
where the functional derivative $\frac{\delta \ell}{\delta u} \in\mathfrak{g}  ^\ast  $ is defined as 
\begin{equation}\label{definition_fd}
\left\langle \frac{\delta \ell}{\delta u}, \delta u \right\rangle _\mathfrak{g}  := \left.\frac{d}{d\varepsilon}\right|_{\varepsilon=0} \ell(u+ \varepsilon \delta u),
\end{equation} 
and $\operatorname{ad}^*_u: \mathfrak{g}^\ast \rightarrow \mathfrak{g}  ^\ast $ denotes the coadjoint operator defined by $ \left\langle \operatorname{ad}^*_u m, v \right\rangle _ \mathfrak{g}  :=\left\langle m, [u,v] \right\rangle_\mathfrak{g}$.

\subsection{The stochastic Hamilton--Clebsch variational principle \cite{holm2015variational}}

Let us now consider, as before, the configuration Lie group $G=\operatorname{Diff}_{\rm vol}(\mathcal{D})$ of the incompressible fluid, and a right $G$-invariant Lagrangian $L:TG \rightarrow\mathbb{R}$ with reduced Lagrangian $\ell:\mathfrak{g}\rightarrow\mathbb{R}$. We shall assume in addition that $G$ acts on the right on a vector space $V$, usually given by a space of tensor fields on $\mathcal{D}$, and we denote by $\pounds_uq\in V$ the infinitesimal generator of this action, for $u\in\mathfrak{g}$.
Given the $N$ time independent divergence free vector fields $ \xi _i(x)$, $i=1,...,N$, in \eqref{Lag-stoch-process}, the stochastic Hamilton--Clebsch constrained variational principle in \cite{holm2015variational} is formally written as
\begin{equation}\label{Clebsch_VP} 
\delta\int_0^T \Big[\ell(u)dt + \left\langle p, {\dd}q + \pounds_{{\dd}x_t}q\right\rangle_V\Big]=0,
\end{equation} 
with respect to variations $ \delta u$, $ \delta q$, $ \delta p$, for $\delta q$ vanishing at $t=0,T$, and where $\left\langle\,\cdot \,, \,\cdot\,\right\rangle _V$ denotes the duality pairing between $V$ and its dual space $V^*$. Here ${\dd}x_t$ is defined as in \eqref{Lag-stoch-process}, and may be rewritten equivalently in Eulerian form as
\begin{equation}\label{def_dx} 
{\dd}x_t:=u(t,x) dt+ \sum_{i=1}^N\xi_i(x)\circ dW_i(t)\,.
\end{equation} 
The variations in \eqref{Clebsch_VP} with respect to $\delta u$, $\delta p$ and $\delta q$ yield, respectively, the conditions
\begin{equation}\label{conditions} 
\frac{\delta \ell}{\delta u}= p \diamond q,\qquad {\dd}q+ \pounds _{{\dd}x_t}q=0
, \qquad {\dd}p- \pounds ^\mathsf{T}_{{\dd}x_t}p=0,
\end{equation} 
where $ p \diamond q \in \mathfrak{g}  ^\ast$ and $\pounds ^\mathsf{T}_up\in V^\ast $ are defined as 
\begin{equation}\label{definition_diamond}
\left\langle p \diamond q, u \right\rangle _ \mathfrak{g}  = \left\langle p,\pounds _ u q \right\rangle _V , \qquad \left\langle\pounds _u^\mathsf{T} p, q \right\rangle _V= \left\langle p,\pounds _u q \right\rangle _V ,
\end{equation} 
for $q \in V$, $p \in V ^\ast $, and $u, \delta u \in \mathfrak{g}  $.
The stationarity conditions \eqref{conditions} imply the following stochastic Euler--Poincar\'e equation:
\begin{equation}\label{Stoch_EP}
\dd \frac{\delta \ell}{\delta u}+ \operatorname{ad}^* _{{\dd}x_t} \frac{\delta \ell}{\delta u} =0,
\end{equation}
where $\operatorname{ad}^*_u: \mathfrak{g}^\ast \rightarrow \mathfrak{g}  ^\ast $ denotes as before the coadjoint operator and $ dx_t$ is given in \eqref{def_dx}.

The notations used in \eqref{Stoch_EP} are general enough to make this equation valid for any Lie group $G$ and Lagrangian $ \ell: \mathfrak{g}  \rightarrow \mathbb{R} $. For example, see \cite{ArHoCa2017} for a parallel treatment for the rigid body and the group $SO(3)$, as well as for the heavy top, which involves advected quantities arising from symmetry breaking from $SO(3)$ to $SO(2)$. See \cite{CrHoRa2017} for more discussions of the general case of stochastic Euler--Poincar\'e equations in finite dimensions. 

\paragraph{An example: Euler's fluid equations in 3D and 2D.}
Upon choosing for $ \mathfrak{g}  ^\ast $ the space of divergence free vector fields on $ \mathcal{D} $ parallel to the boundary $\partial \mathcal{D}$, i.e., $ \mathfrak{g}  ^\ast = \mathfrak{g}  $, and the duality pairing 
\[
\left\langle m,u \right\rangle _ \mathfrak{g}  = \int_ \mathcal{D} m(x) \!\cdot\! u(x) \,d^nx\,,
\]
one obtains the coadjoint operator as $\operatorname{ad}^*_u m= \mathbb{P}  ( u\cdot \nabla m+ \nabla u^\mathsf{T} \cdot m)$, where $ \mathbb{P}$ is the Hodge projection onto divergence free vector fields parallel to the boundary. With the Lagrangian \eqref{Clebsch_VP}, the stochastic Euler--Poincar\'e equation \eqref{Stoch_EP} becomes, in 3D,
\begin{equation}\label{stoch_3D_Euler} 
{\dd} u+ \mathbb{P}  ( u \cdot \nabla u) dt+\sum_{i=1}^N \mathbb{P}  ( \operatorname{curl} u \times \xi_i  ) \circ dW _i (t) =0\,.
\end{equation} 
{
Equation \eqref{stoch_3D_Euler} can be written equivalently in vorticity form as
\begin{equation}\label{stoch_vort_eqn} 
{\dd}\omega+({\dd}x_t\cdot\nabla)\omega - (\omega\cdot\nabla){\dd}x_t=0,
\end{equation} 
where $\omega={\rm curl}\, u$ is the vorticity and the stochastic vector field ${dx_t}$ is given in equation \eqref{def_dx}. 
}

When $\mathcal{D}$ is a simply connected bounded domain in $\mathbb{R}^2$, a divergence free vector field $u$ has a unique associated stream function $\psi$ such that $u=\mathbf{\hat{z}}\times \nabla \psi$ and $\psi|_{\partial\mathcal{D}}=0$, where $\mathbf{\hat{z}}$ is the unit vector of the $z$-axis pointing upward. The dual space $ \mathfrak{g}  ^\ast $ is identified with the space of absolute vorticities $ \varpi $ on $\mathcal{D}$, via the duality pairing
\[
\left\langle \varpi  , \psi \right\rangle _ \mathfrak{g}  = \int_ \mathcal{D} \varpi (x)\psi (x) d ^2 x.
\]
The absolute vorticity $\varpi$ is related to the total fluid momentum $m$ as $ \varpi = \operatorname{curl} m \cdot \mathbf{\hat{z}} $, where $\mathbf{\hat{z}}$ is the vertical unit vector. For instance, for the Euler equation, the absolute vorticity coincides with the vorticity $ \omega = \operatorname{curl} u \cdot \mathbf{\hat{z}}=\Delta\psi$, whereas for the rotating Euler equation, we have $ \varpi  = \operatorname{curl} u \cdot \mathbf{\hat{z}}+f= \omega +f$, where $f$ is the Coriolis parameter, which depends on latitude for motion on the Earth. On non simply connected domains, with $K$ holes with smooth boundary $\partial \mathcal{D}_k$, $k=1,...,K$, the stream function associated to a given velocity field $u$ is determined by the condition $\psi|_{\partial\mathcal{D}_0}=0$ and satisfies $\psi|_{\partial\mathcal{D}_k}=c_k$, where $\partial\mathcal{D}_0$ is the outer boundary and $c_k$, $k=1,...,K$ are constant. In this case, the dual space has to be augmented with the circulations numbers $\Gamma_k$, $k=1,...,K$ around each hole, see  \cite{marsden1983coadjoint}.

In 2D, the stochastic Euler equation \eqref{Stoch_EP} becomes
\begin{equation}\label{stoch_2D_Euler} 
{\dd}\omega + \{ \omega , \psi \} dt+ \sum_{i=1}^ N\{ \omega , \zeta_i \}\circ dW _i (t) =0,
\end{equation} 
where, for two functions $f,g$ on $\mathcal{D} $, the function $\{f,g\}$ is the Jacobian defined by $ \{f,g\}:= \partial _{x _1 } f\partial _{x_2} g - \partial _{x _2 } f\partial _{x_1} g  $, with $x=(x_1 , x _2 )$. In \eqref{stoch_2D_Euler}, $ \psi (t,x)$ is the stream function of the fluid velocity $u(t,x)=\mathbf{\hat{z}}\times \nabla \psi(t,x)$, the variable $ \omega (t,x)= \Delta \psi (t,x)$ is its vorticity, and the functions $ \zeta_i (x)$ are the stream functions of the divergence free vector fields $ \xi _i=\mathbf{\hat{z}}\times \nabla \zeta_i$, where we recall that $\zeta_i$ is zero on $\partial \mathcal{D}_0$ and constant on $\partial \mathcal{D}_k$, $k=1,..,K$. 
The deterministic Euler equations are recovered in \eqref{stoch_3D_Euler} and \eqref{stoch_2D_Euler} when $ \xi _i =0$, for all $=1,...,N$.

\paragraph{Remark.} In the present paper, we shall consider stochastic variational principles in infinite dimensions only in a formal sense, for the purpose of modelling time-dependent spatial correlations. Some of the fundamental questions in analysis for the stochastic 3D Euler fluid model have been answered in \cite{CrFlHo2017}, who proved local in time existence, uniqueness and well posedness of solutions in regular spaces, as well as a Beale-Kato-Majda blow-up criterion for these equations. These are precisely the same analytical properties as for the deterministic 3D Euler fluid equations. Thus, in this case, introducing stochasticity that preserved the geometric properties of the Euler fluid equations also preserved their analytical properties.

\subsection{The stochastic Hamilton--Pontryagin variational principle \cite{gaybalmaz2017geometric}}

Knowing that the deterministic Euler fluid equations in the Lagrangian fluid description arise from the Hamilton principle \eqref{HP}, or the Hamilton-Pontryagin principle \eqref{Ham_Pont}, for the right invariant Lagrangian $L:T G\rightarrow \mathbb{R}$ given by the kinetic energy, 
we expect the stochastic Euler fluid equations \eqref{Stoch_EP} to arise, in the Lagrangian description, via a stochastic extension of these principles. This is indeed the case if one proceeds formally here and below, by considering the \textit{stochastic Hamilton-Pontryagin  (SHP) principle}
\begin{equation}\label{SVP}
\delta \int_0^T \Big[L(g_t, v)dt+ \big\langle \pi , {\dd}g_t- vdt - \sum_{i=1}^N\xi_ig_t\circ d W _i(t) \big\rangle \Big]=0\,,
\end{equation}
with respect to variations $ \delta g_t$, $ \delta v$, $ \delta \pi $, for $\delta g_t$ vanishing at $t=0,T$. The variables $v$ and $ \pi $ are, respectively, the material fluid velocity and material fluid momentum. As before, the angle brackets $ \left\langle\,\cdot \,, \,\cdot\,\right\rangle $ denote the pairing between elements in $T_g^\ast G$, and $T_g G$, the cotangent and tangent space to $G$ at $g$. The notation $ \xi _i g_t$ indicates the composition of the vector field $ \xi _i $ on the right by the diffeomorphic flow $g_t$.  

Stochastic Hamilton-Pontryagin principles (SHPs) have been considered for finite dimensions in \cite{bourabee2009stochastic}. 
The SHP was considered in infinite dimensions for the first time in \cite{gaybalmaz2017geometric}, where it was shown to afford a systematic derivation of the stochastic equations that preserves their deterministic mathematical properties, both geometrical and analytical.

Note that \eqref{SVP} imposes the stochastic process \eqref{Lag-stoch-process} as a constraint on the variations by using the Lagrange multiplier $ \pi $. From the $G$-invariance of both the Lagrangian and the constraint, this principle can be equivalently written formally in the reduced Eulerian description as
\begin{equation}\label{Reduced_SVP}
\delta \int_0^T \Big[\ell(u)dt
+ \big\langle m, {\dd}g_t\,g_t^{-1}- udt 
- \sum_{i=1}^N\xi_i\circ d W _i(t) \big\rangle _ \mathfrak{g}  \Big]=0\,,
\end{equation}
with respect to variations $ \delta u, \delta g, \delta m $, and where $u= v g_t ^{-1} \in \mathfrak{g}   $, $ m = \pi g_t ^{-1} \in \mathfrak{g}  ^\ast  $.
This is the reduced stochastic Hamilton-Pontryagin (RSHP) principle found in \cite{gaybalmaz2017geometric}. It is clearly a stochastic extension of the reduced Hamilton-Pontryagin principle \eqref{Ham_Pont_Eulerian}. Its stationarity conditions are
\[
{\dd} g_t g_t^{-1}= {u}dt + \sum_{i=1}^N\xi_i\circ d W _i(t),\qquad \frac{\delta \ell}{\delta {u}}=m,\qquad {\dd} m+ \operatorname{ad}^*_{{\dd} g_t g_t^{-1}}m=0,
\]
which can directly be compared to their deterministic counterparts obtained in \eqref{Stat_cond_Eulerian}.

One then directly checks that the stochastic variational principle \eqref{Reduced_SVP} also yields the stochastic equation \eqref{Stoch_EP}. Thus, the two variational principles \eqref{Clebsch_VP} and \eqref{Reduced_SVP} both yield the same stochastic equations. Moreover, in absence of stochasticity, equation \eqref{SVP} recovers the Hamilton-Pontryagin principle for Lagrangian mechanics, see \cite{yoshimura2006dirac}.

\paragraph{Remark.} The RSHP principle in \eqref{Reduced_SVP} has several interesting properties: (i) it allows a formulation of reduction by symmetry in the stochastic context; (ii) it  does not need the introduction of the extra advected quantities $q,p$; and (iii) it does not restrict the values of the Eulerian fluid momentum $m \in \mathfrak{g}  ^\ast $ to be of the form, $m=p\,\diamond \,q$. 
Finally, we note that the SHP principle \eqref{SVP} is not restricted to configuration manifolds which are Lie groups. SHP can be written for Lagrangian systems on a smooth manifold $Q$ as
\begin{equation}\label{SVP_manifold}
\delta \int_0^T \Big[L(q_t, v)dt+ \big\langle \pi , {\dd}q_t- vdt - \sum_{i=1}^NX_i(q)\circ d W _i(t) \big\rangle \Big]=0\,,
\end{equation}
for variations $ \delta q_t$, $ \delta v$, $ \delta \pi $, with $\delta q_t$ vanishing at $t=0,T$, and for given vector fields $X_i$ on $Q$, $i=1,...,N$. When $Q=G$ and the vector fields $X_i$ are right $G$-invariant, \eqref{SVP} is recovered.

\section{Stochastic Hamiltonian formulations} The SHP principle \eqref{SVP} can be equivalently written as 
\begin{equation}\label{Ham_SVP_1}
\delta \int _0 ^T  \Big[L(g_t, v)+ \big\langle  \pi , {\dd}g_t- vdt \big\rangle- \sum_{i=1}^NH_i(g_t,\pi; \xi _i  )\circ d W _i(t) \Big] =0\,,
\end{equation}
for the $G$-invariant functions $H_i(\_\,,\_\,; \xi _i ):T^*G \rightarrow \mathbb{R}$ defined by
\begin{equation}\label{H_i}
H_i(g_t, \pi ; \xi _i ):= \langle\pi , \xi_i g_t \rangle= \left\langle \pi g_t ^{-1}  ,\xi_i\right\rangle_ \mathfrak{g}  , \;\; i=1,...,N .
\end{equation} 
The form of the variational principle in \eqref{Ham_SVP_1} allows for the derivation of other stochastic geometric models by appropriate choices of the stochastic Hamiltonians $ H _i $ and their symmetries, see  \cite{gaybalmaz2017geometric}, not necessarily of the form \eqref{H_i}.  

The variational principle in \eqref{Ham_SVP_1} yields the following stochastic extension of the Euler-Lagrange equations with Lagrangian $L$:
\begin{equation}\label{Lag_form} 
{\dd} \frac{\partial L}{\partial v}=\frac{\partial L}{\partial g_t}dt- \sum_{i=1}^N\frac{\partial H _i }{\partial g_t} \circ dW _i (t) =0, 
\quad  {\dd}g_t=v dt+\frac{\partial H_i }{\partial \pi }\circ dW _i (t) , \quad \pi = \frac{\partial L}{\partial v}\,.
\end{equation} 
This is the Lagrangian description of the stochastic equations \eqref{Stoch_EP}.
Denoting by $H: T^*G \rightarrow \mathbb{R}  $,  the Hamiltonian associated to $L$ by the Legendre transform, we can rewrite these equations in stochastic Hamiltonian form
\begin{equation}\label{Ham_form} 
{\dd}g_t = \frac{\partial ({\dd}H) }{\partial \pi } = \frac{\partial H}{\partial \pi } dt+  \sum_{i=1}^N \frac{\partial H _i }{\partial\pi }\circ d W _i (t)  
, \quad 
{\dd}\pi = -\frac{\partial ({\dd}H) }{\partial \pi } = -   \frac{\partial H}{\partial g_t } dt- \sum_{i=1}^N  \frac{\partial H _i }{ \partial g_t }\circ d W _i (t)\,,
\end{equation} 
with
\begin{equation}\label{Ham-stoch} 
{\dd}H := H(g_t, \pi )\,dt + \sum_{i=1}^N H_i(g_t, \pi ; \xi _i )  \circ d W _i (t)\,.
\end{equation} 
Consequently, we can call the functions $H _i $ the \textit{stochastic Hamiltonians}. Stochastic Hamiltonian systems of the form \eqref{Ham_form} were first developed in \cite{bismut1982mecanique}. 

These equations can be written in terms of the \textit{canonical Poisson bracket} $\{\,\cdot\,,\,\cdot\,\}_{\rm can}$ on $T^*G$ as
\begin{equation}\label{Poisson_form} 
{\dd}F = \{ F, {\dd}H\} _{\rm can} = \{F,H\} _{\rm can}dt+\sum_{i=1}^N \{F, H_i \}_{\rm can}\circ d W _i  (t) \,,
\end{equation} 
for arbitrary functionals $F=F(g, \pi ): T^*G \rightarrow \mathbb{R}  $. 

Consistently with this observation, the stochastic Euler fluid equation \eqref{Stoch_EP} can also be written in Hamiltonian form as
\begin{equation}\label{LP_stoch}
{\dd} m + \operatorname{ad}^*_{ \frac{\delta ({\dd}h)}{\delta m }}m
=
{\dd} m + \operatorname{ad}^*_{ \frac{\delta h}{\delta m }}m \,dt+ \sum_{i=1}^N \operatorname{ad}^*_{ \frac{\delta h _i }{\delta m }  } m \circ dW _i (t) =0\,,
\end{equation} 
where $h: \mathfrak{g}  ^\ast \rightarrow \mathbb{R}  $ and $h _i :\mathfrak{g}  ^\ast \rightarrow \mathbb{R}$ are the reduced Hamiltonians associated to $H$ and $H _i $ in \eqref{H_i}, i.e., $H(g_t, \pi )= h( \pi g_t ^{-1} )$ and $H _i (g_t,\pi ;\xi _i )= h_i (\pi  g_t^{-1} )$, and
\begin{equation}\label{ham-stoch} 
{\dd}h :=  h(m)\,dt + \sum_{i=1}^N h_i(m ; \xi _i )  \circ d W _i (t)\,.
\end{equation} 
Upon comparison with equation \eqref{Stoch_EP}, we find
\begin{equation}\label{ed_hi}
h(m,t)= \int_ \mathcal{D} \frac{1}{2} |m(x,t)| ^2 d^nx\quad\text{and}\quad  h _i ( m )=  \left\langle m , \xi_i\right\rangle_ \mathfrak{g} =\int_ \mathcal{D} m(x,t)\!\cdot \! \xi _i (x) \,d^nx.
\end{equation}
The expression \eqref{LP_stoch} is the reduced (or Euler--Poincar\'e) formulation of the Hamiltonian formulation \eqref{Ham_form}. 

In terms of the Lie-Poisson bracket $\{\,\cdot \,,\,\cdot\,\}_{\rm LP}$ on $ \mathfrak{g}  ^\ast $, given by
\[
\big\{f,({\dd}h)\big\}_{\rm LP}( m )= \left\langle m , \left[\frac{\delta f}{\delta m }, \frac{\delta {(\dd}h)}{\delta m }\right] \right\rangle_ \mathfrak{g}  ,
\]
equation \eqref{Stoch_EP} and hence \eqref{LP_stoch} can be formulated in the Stratonovich-Lie-Poisson form
\begin{equation}\label{LP_form} 
{\dd}f = \big{\{f,(\dd}h)\big\} _{\rm LP} = \{f,h\} _{\rm LP}dt+\sum_{i=1}^N \{f, h_i \}_{\rm LP}\circ d W _i(t)  \,,
\end{equation}
for arbitrary functions $f: \mathfrak{g}  ^\ast \rightarrow \mathbb{R}  $.
This is the reduced form of \eqref{Poisson_form}. The Poisson bracket formulation is especially useful to convert the equations into their It\^o version, whose It\^o-Lie-Poisson form is 
\[
{\dd}f = \left(\{f,h\} _{\rm LP}+ \frac{1}{2}\sum_{i=1}^N\{h_i,\{h_i,f\}\} _{\rm LP} \right)dt+\sum_{i=1}^N \{f, h_i \}_{\rm LP}\cdot d W _i(t)  \,.
\]

For example, the stochastic 2D Euler equations \eqref{stoch_2D_Euler}, on a simply connected domain with boundary, can be written in the Stratonovich-Lie-Poisson form \eqref{LP_form} with  the Lie-Poisson bracket written on the space of vorticities as \cite{marsden1983coadjoint}
\begin{equation}\label{LP_bracket} 
\{f,g\}_{\rm LP}( \omega )= \int_ \mathcal{D}  \omega \left\{ \frac{\delta f}{\delta \omega },\frac{\delta g}{\delta \omega }   \right\} d^2 x
\end{equation} 
and with the Hamiltonian and stochastic Hamiltonians
\[
h (\omega )= -\frac12\int _ \mathcal{D} \omega (x,t) \,\psi (x,t) \, d ^2 x
\,,\quad\hbox{and}\quad
h _i ( \omega )=-\int _ \mathcal{D} \omega (x,t) \,\zeta_i (x) \, d ^2 x\,,
\]
where $\xi_i(x)=\mathbf{\hat{z}}\times\nabla \zeta_i(x)$, with $\zeta|_{\partial\mathcal{D}}=0$, i.e.,  the $\zeta_i(x)$ are stream functions for the $\xi_i(x)$.
In domains which are not simply connected, the dual space includes the circulations $(\omega, \Gamma_1,...,\Gamma_K)$ in case of $K$ islands, see \cite{marsden1983coadjoint}, with Hamiltonian $h (\omega ,\Gamma_1,...,\Gamma_K)= -\frac12\int _ \mathcal{D} \omega (x,t) \,\psi (x,t) \, d ^2 x+\sum_k c_k \Gamma_k$, where $c_k=\psi|_{\partial\mathcal{D}_k}$.

\paragraph{Remark.} As mentioned above after equation \eqref{SVP_manifold}, the variational principle \eqref{Ham_SVP_1} admits a natural extension to general configuration manifolds $Q$ as
\begin{equation}\label{Ham_SVP_1_manifold}
\delta \int _0 ^T  \Big[L(q_t, v)+ \big\langle  \pi , {\dd}q_t- vdt \big\rangle- \sum_{i=1}^NH_i(q_t,\pi )\circ d W _i(t) \Big] =0\,,
\end{equation}
for given Hamiltonians $H_i:T^*Q\rightarrow \mathbb{R}$, $i=1,...,N$. When the Lagrangian is hyperregular, the principle can be reformulated exclusively on the Hamiltonian side as
\[
\delta \int _0 ^T  \Big[\big\langle  \pi , {\dd}q_t\big\rangle- H(q_t,\pi)dt -  \sum_{i=1}^NH_i(q_t,\pi )\circ d W _i(t) \Big] =0\,,
\]
for variations $\delta q_t$ and $\delta p$, with $\delta q_t$ vanishing at $t=0,T$.

\section{Example: Hamiltonian equations of motion for a multi-layer fluid}

\subsection{A deterministic $N$-layer quasigeostrophic (NLQG) fluid}

Consider a stratified fluid of $N$ superimposed layers of constant densities $\rho_1 < \dots<\rho_N$; the layers being stacked according to increasing density, such that the density of the upper layer is $\rho_1$. The quasigeostrophic (QG) approximation assumes that the velocity field is constant in the vertical direction and that in the horizontal direction the motion obeys a system of coupled incompressible shallow water equations. We shall denote by $\mathbf{u}_i = (- \,\partial_y\psi_i, \partial_x\psi_i) = \mathbf{\hat{z}}\times\nabla \psi_i$ the velocity field of the $i^{th}$ layer, where $\psi_i$ is its stream function, and the layers are numbered from the top to the bottom. We define the generalised total vorticity of the $i^{th}$ layer as
\begin{equation}\label{omsubi}
\omega_i = q_i + f_i = \Delta  \psi_i + \alpha_i \sum_{j=1}^N T_{ij}\psi_j + f_i
=: \sum_{j=1}^N E_{ij}\psi_j +  f_i
\,,\qquad
i=1,\dots,N,
\end{equation}
where the generalised total vorticity  is defined as $\omega_i = q_i + f_i $, the  elliptic operator $E_{ij}$ defines the layer vorticity,
\[q_i = \sum_{j=1}^N E_{ij}\psi_j:= \Delta \psi_i + \alpha_i \sum_{j=1}^N T_{ij}\psi_j\,,\] 
and the constant parameters $\alpha_i $, $f_i $, $f_0$, $\beta$, $f_N$ are
\begin{align}\label{paramdefs}
\begin{split}
\alpha_i &= (f_0^2/g)\big((\rho_{i+1}-\rho_i)/\rho_0\big)D_i
\,,\qquad i=1,\dots,N,
\\
f_i &= f_0 + \beta y
\,,\qquad  i=1,\dots,N-1,
\\
f_N &= f_0 + \beta y + f_0 d(y)/D_N,
\\
f_0 &= 2\Omega \sin(\phi_0)
\,,\qquad 
\beta = 2\Omega \cos(\phi_0)/R\,,
\end{split}
\end{align}
where $g$ is the gravitational acceleration, $\rho_0 = (1/N)(\rho_1 + \dots + \rho_N)$ is the mean density, $D_i$ is the mean thickness of the $i^{th}$ layer, $R$ is the Earth's radius, $\Omega$ is the Earth's angular velocity, $\phi_0$ is the reference latitude, and $d(y)$ is the shape of the bottom. The $N \times N$ symmetric tri-diagonal matrix $T_{ij}$ represents the second-order difference operator,
\begin{equation}\label{2ndDiffTop}
\sum_{j=1}^N T_{ij}\psi_j  = (\psi_{i-1} - \psi_i) - (\psi_i - \psi_{i+1})\,,
\end{equation}
so that 
\begin{equation}\label{2ndDiffT}
T_{ij} = 
\begin{bmatrix}
-1 & 1 & 0 &  0 & \dots &\dots &  0
\\
1 &  -2  & 1 & 0 & \dots & \dots & 0 
\\ 0 & 1  & \dots &  \dots & \dots & 1 & 0 
\\ 0 &\dots &  \dots & 0 &    1 & -2  & 1 
\\ 0 &\dots &  \dots & 0 &  0 & 1 & -1 
\end{bmatrix}
\,,\qquad
i,j=1,\dots,N.
\end{equation}
With these standard notations, the motion of the NLQG fluid is given by
\begin{equation}\label{NlayerVortDyn}
\partial_t q_i  = \Big\{ \omega_i ,\,\psi_i \Big\}_{xy}
=
-\, \mathbf{\hat{z}} \times \nabla \psi_i  \cdot \nabla \omega_i
=
-\,
\mathbf{u}_i \cdot \nabla \omega_i
\,,\qquad
i =1,\dots,N,
\end{equation}
where $\mathbf{\hat{z}}$ is the vertical unit vector, $\mathbf{u}_i = 
\mathbf{\hat{z}} \times \nabla \psi_i  $ is the horizontal flow velocity in the $i^{th}$ layer, and the brackets in 
\begin{equation}\label{canPoissonBrkt}
\{\omega,\psi\}=J(\omega,\psi)=\omega_x\psi_y-\omega_y\psi_x
= \mathbf{\hat{z}}\cdot \nabla \omega \times \nabla \psi 
\end{equation}
denote the usual $xy$ canonical Poisson bracket in $\mathbb{R}^2$.  As before, the boundary conditions for the stream functions in a compact domain $\mathcal{D}\subset\mathbb{R}^2$ with $K$ holes, are $\psi_j|_{\partial \mathcal{D}_0}=0$ and $\psi_j|_{\partial\mathcal{D}_k} = constant$, $k=1,...,K$, whereas in the entire $\mathbb{R}^2$ they are
$\lim_{(x, y)\to±\infty} \nabla\psi_j=0$.
The space of variables with canonical Poisson bracket in \eqref{canPoissonBrkt} consists of $N$-tuples $(q_1,\dots, q_N)$ of real-valued functions on $\mathcal{D}$ (the ``generalized vorticities'') with certain smoothness properties that guarantee that solutions are at least of class $C^1$.
The Hamiltonian for the $N$-layer vorticity dynamics in \eqref{NlayerVortDyn} is the total energy
\begin{equation}\label{NlayerVortDyn-erg}
H(q_1,\dots, q_N) = \frac12\int_\mathcal{D}
\Big[\sum_{i=1}^N \frac{1}{\alpha_i} |\nabla\psi_i|^2
+
\sum_{i=1}^{N-1} (\psi_{i+1}-\psi_i)^2\Big]dx\,dy
\,,\qquad
i =1,\dots,N,
\end{equation}
with stream function $\psi_i$ determined from vorticity $\omega_i$ by solving the elliptic equation \eqref{omsubi} for $q_i=\omega_i-f_i$ with 
\begin{equation}\label{elliptic-op}
q_i = \sum_{j=1}^N E_{ij}\psi_j\,,
\end{equation}
for the boundary conditions discussed above. Hence, we find that 
\begin{equation}\label{NlayerVortDyn-erq}
H(q_1,\dots, q_N) 
= -\frac12\int_\mathcal{D}\sum_{i,j=1}^N \psi_iE_{ij}\psi_j dx\,dy
= -\frac12\int_\mathcal{D}\sum_{i,j=1}^N q_i E^{-1}_{ij}*q_j dx\,dy
= -\frac12\int_\mathcal{D}\sum_{i=1}^N q_i \psi_i dx\,dy\,,
\end{equation}
where $E^{-1}_{ij}*q_j = \psi_i$ denotes convolution with the Greens function $E^{-1}_{ij}$ for the symmetric elliptic operator $E_{ij}$. The relation \eqref{NlayerVortDyn-erq} means that $\delta H/\delta q_i = \psi_i$ for the variational derivative of the Hamiltonian functional $H$ with respect to the function $q_j$.
As before, if the domain $\mathcal{D}$ is not simply connected, the dual space will include the circulations around each boundary, for each layer.

\paragraph{\bf Lie--Poisson bracket.}
Equations \eqref{NlayerVortDyn} are Hamiltonian with respect to the Lie--Poisson bracket given by
\begin{equation}\label{VortLie-PoissonBrkt}
\{F,H\}(q_1,\dots, q_N)
=
\sum_{i=1}^N \int_\mathcal{D} (q_i + f_i(x))
\left\{\frac{\delta F}{\delta q_i},\,\frac{\delta H}{\delta q_i}\right\}_{xy}dx\,dy\,,
\end{equation}
provided the domain of flow $\mathcal{D}$ is simply connected.%
\footnote{If the domain $\mathcal{D}$ is not simply connected, then variational derivatives such as $\delta H/\delta q_i$ must be interpreted with care, because in that case the boundary conditions on $\psi_i$ will come into play \cite{McWilliams1977}.}

The motion equations \eqref{NlayerVortDyn} for $q_i$ now follow from the Lie--Poisson bracket \eqref{VortLie-PoissonBrkt} after an integration by parts to write it equivalently as
\begin{equation}\label{VortLie-PoissonBrkt2}
\frac{dF}{dt} = \{F,H\}(q_1,\dots, q_N)
=
-\sum_{i=1}^N \int_\mathcal{D} \frac{\delta F}{\delta q_i}
\left\{q_i + f_i(x) ,\,\frac{\delta H}{\delta q_i}\right\}_{xy}dx\,dy
\,,
\end{equation}
and recalling that $\delta H/\delta q_i =-E^{-1}_{ij}*q_j=- \,\psi_i$, $i=1,2,\dots,N$, so that equations \eqref{NlayerVortDyn} follow. 

\paragraph{\bf Constants of motion.}
According to equations \eqref{NlayerVortDyn}, the material time derivative of $\omega_i(t, x, y)$ vanishes along the flow lines of the divergence-free horizontal velocity $\mathbf{u}_i = \mathbf{\hat{z}}\times\nabla \psi_i   $. Consequently, for every differentiable function $\Phi_i: \mathbb{R}\to\mathbb{R}$ the functional
\begin{equation}\label{Casimirs}
C_{\Phi_i}(\omega_i)
=
\int_\mathcal{D} \Phi_i (\omega_i)\,dx\,dy
\end{equation}
is a conserved quantity for the system \eqref{NlayerVortDyn} for $i =1,\dots,N$, provided the integrals exist. By Kelvin's circulation theorem, the following integrals over an advected domain $S(t)$ in the plane are also conserved,
\begin{equation}\label{KelvinThm}
I_i(t) 
=
\int_{S(t)} \omega_i \,dx\,dy
= 
\int_{\partial S(t)}  \nabla \psi_i \cdot \mathbf{\hat{n}} \,ds
\,,
\end{equation}
where $\mathbf{\hat{n}}$ is the horizontal outward unit normal and $ds$ is the arclength parameter of the closed curve $\partial S(t)$ bounding the domain $S(t)$ moving with the flow.

\subsection{Hamiltonian formulation for the stochastic NLQG fluid}

Having understood the geometric structure (Lie--Poisson bracket, constants of motion and Kelvin circulation theorem) for the deterministic case, we can introduce the stochastic versions of equations \eqref{NlayerVortDyn} by simply making the Hamiltonian stochastic while preserving the previous geometric structure, as done in the previous section. Namely, we choose
\begin{equation}\label{Ham-stoch}
\dd h = H(\{q\})dt + \int_D \sum_{i=1}^N\sum_{k=1}^K q_i (t,x,y)\zeta^k_i(x,y) \circ dW_k(t)\,dx\,dy 
\,,
\end{equation}
where the $\zeta^k_i(x,y)$, $k =1,\dots,K$ represent the correlations of the Stratonovich noise we have introduced in \eqref{Ham-stoch}. 

For this stochastic Hamiltonian, the Lie--Poisson bracket \eqref{VortLie-PoissonBrkt}  leads to the following stochastic process for the transport of the $N$-layer generalised vorticies,
\begin{equation}\label{NlayerVortDyn-stoch}
\dd q_i  = \Big\{ \omega_i ,\,\dd \psi \Big\}_{xy}
=
J\big(\omega_i ,\,\dd \psi \big)
=
\nabla (\dd \psi_i ) \times \mathbf{\hat{z}}\cdot \nabla \omega_i
=
-\,
\dd \mathbf{u}_i \cdot \nabla \omega_i
\,,\qquad
i =1,\dots,N,
\end{equation}
where we have defined the stochastic transport velocity in the $i^{th}$ layer
\begin{equation}\label{Nlayer-stoch-vel}
\dd \mathbf{u}_i :=
\mathbf{\hat{z}} \times \nabla (\dd \psi_i )  
\,,\qquad
i =1,\dots,N,
\end{equation}
in terms of its stochastic stream function 
\begin{equation}\label{Nlayer-stoch-dpsi}
\dd \psi_i  := \psi_i \,dt + \sum_{k=1}^K \zeta^k_i(x,y) \circ dW_k(t)
= \frac{\delta(\dd h)}{\delta q_i}
\,,\qquad
i =1,\dots,N,
\end{equation}
determined from the variational derivative of the stochastic Hamiltonian in \eqref{Ham-stoch} with respect to the generalised vorticity $q_i$ in the $i^{th}$ layer. 

\paragraph{\bf Constants of motion.} The constants of motion $C_{\Phi_i}$ in \eqref{Casimirs} and the Kelvin circulation theorem for the integrals $I_i$ in \eqref{KelvinThm} persist for the stochastic generalised vorticity equations in \eqref{NlayerVortDyn-stoch}. This is because both of these properties follow from the Lie-Poisson bracket in \eqref{VortLie-PoissonBrkt}. However, the stochastic Hamiltonian in \eqref{Ham-stoch} is not conserved, since it depends explicitly on time, $t$, through its Stratonovich noise term.

\section{Outlook}

This brief note has reviewed only a small fraction of what has been happening in stochastic geometric mechanics recently, in the hopes that Juergen would become interested in it.  The idea of structure preserving stochasticity is very powerful. See, for example, Albeverio et al. \cite{Albeverio-etal-CIB-SGM} for some other perspectives. 

\subsection*{Acknowledgements} 
The authors are grateful for many stimulating and thoughtful discussions about Stochastic Geometric Mechanics with S. Albeverio, A. Arnaudon, A. B. Cruzeiro, D. O. Crisan, C. J. Cotter, F. Flandoli, T. S. Ratiu and T. Tyranowski. FGB is partially supported by the ANR project GEOMFLUID, ANR-14-CE23-0002-01; DDH is grateful for partial support by the EPSRC Standard Grant EP/N023781/1.

%


\bibliographystyle{alpha}
\bibliography{refs,sgm-2015}

\end{document}